\documentclass[%
superscriptaddress,
 amsmath,amssymb,
 aps,twocolumn,
floatfix,
]{revtex4-1}

\usepackage{eurosym}
\usepackage{graphicx}
\usepackage{fancyhdr}
\usepackage{dcolumn}
\usepackage{bm}
\usepackage[colorlinks=True,linkcolor=red,citecolor=blue,urlcolor=blue]{hyperref}

\begin{document}

\title{Quantum computing approach to realistic ESG-friendly stock portfolios}

\author{Francesco Catalano}
\affiliation{Deutsche Börse AG, 60485 Frankfurt am Main, Germany}

\author{Laura Nasello}
\affiliation{Deutsche Börse AG, 60485 Frankfurt am Main, Germany}

\author{Daniel Guterding}
\affiliation{Technische Hochschule Brandenburg, Magdeburger Straße 50, 14770 Brandenburg an der Havel, Germany}
\email{daniel.guterding@th-brandenburg.de}

\date{\today}

\begin{abstract}
Finding an optimal balance between risk and returns in investment portfolios is a central challenge in quantitative finance, often addressed through Markowitz portfolio theory (MPT). While traditional portfolio optimization is carried out in a continuous fashion, as if stocks could be bought in fractional increments, practical implementations often resort to approximations, as fractional stocks are typically not tradeable. While these approximations are effective for large investment budgets, they deteriorate as budgets decrease. To alleviate this issue, a discrete Markowitz portfolio theory (DMPT) with finite budgets and integer stock weights can be formulated, but results in a non-polynomial (NP)-hard problem. Recent progress in quantum processing units (QPUs), including quantum annealers, makes solving DMPT problems feasible. Our study explores portfolio optimization on quantum annealers, establishing a mapping between continuous and discrete Markowitz portfolio theories. We find that correctly normalized discrete portfolios converge to continuous solutions as budgets increase. Our DMPT implementation provides efficient frontier solutions, outperforming traditional rounding methods, even for moderate budgets. Responding to the demand for environmentally and socially responsible investments, we enhance our discrete portfolio optimization with ESG (environmental, social, governance) ratings for EURO STOXX 50 index stocks. We introduce a utility function incorporating ESG ratings to balance risk, return, and ESG-friendliness, and discuss implications for ESG-aware investors.
\end{abstract}

\maketitle

\section{Introduction}
Finding an optimal balance between risk and return of an investment is the primary goal for every investor. For investments in securities markets, this problem has been formalized by Markowitz~\citep{Markowitz1952} in the sense that one needs to find optimal weights for each security, so that the portfolio maximizes the return and minimizes the risk within a given universe of considered securities. Mathematically, this amounts to minimizing the utility function $Q_c :\mathbb{R}^k \to \mathbb{R}$ by finding the appropriate weights $\vec{x} \in \mathbb{R}^k$:
\begin{equation}
\label{eq:markowitzutility}
\min \left\{ Q_c(\vec{r}, \Sigma,\phi) \right\} = \min \left\{ \frac{\phi}{2} \vec{x}^T \Sigma \vec{x} - \vec{r}^T \vec{x} \right\} \,.
\end{equation}
Here $\vec{r}$ denotes the expected returns of each portfolio component, $\phi$ controls the level of risk-aversion and $\Sigma \in \mathbb{R}^{k \times k}$ is the asset price correlation matrix. The minimization is subject to the following constraints, which ensure that the entries of $\vec{x}$ can be interpreted as non-negative weights in a long-only portfolio:
\begin{equation}
\begin{split}
\label{eq:markowitzconstraints}
&\vec{x}^T \vec{x} = 1,\\
&x_i \ge 0~~~i \in {1,\dots,k} \,.
\end{split}
\end{equation}
Since the correlation matrix is positive-definite and symmetric, the utility function $Q_c$ is convex, so that a solution to this optimization problem can be found in polynomial time with linear and quadratic algorithms~\citep{Kolm2014}. 

The vector $\vec{x}$ contains non-negative real numbers, which represent the relative weights of capital allocation to the considered assets. These weights can be multiplied by the amount of available capital to obtain the capital allocation to the respective assets. This approach faces problems when implementing such portfolios in a realistic environment, where traded contracts are discrete and securities prices are finite. Hence, the theoretical capital allocation is in general not commensurate with the discrete securities prices. In practice, this challenge is easily overcome by rounding to the nearest multiple of the securities price. The consequences for large portfolios are mild, since the relative weights of asset allocation are hardly changed by the rounding. For small and intermediate portfolios, the rounding may affect the relative weighting significantly and create sub-optimal implementations of originally optimal portfolios.

Discrete extensions of the Markowitz portfolio optimization, where the discreteness of securities contracts is considered from the start, have been studied for a long time, because such discrete portfolios also facilitate the inclusion of further realistic features such as transaction costs or Boolean constraints on stock selection.~\citep{Young1998, Rubio2022, Mugel2022} Intensive studies have been conducted on the problem of transaction costs and the optimal investment trajectory in a multi-period setting. These studies revealed that the discrete Markowitz portfolio theory (DMPT) is a non-polynomial hard problem~\citep{Mansini1999,Coleman2006,Kellerer2000,Jobst2001,Bonami2009}, even if the trajectory problem is only formulated for a single period~\citep{Rosenberg2015}.

The main problem is that the number of possible portfolio compositions grows factorially with the number of assets in the investment universe and the allowed number of assets in the portfolio. If our portfolio may contain $n$ not necessarily different lots out of an investment universe of $k$ different assets, where each asset can be bought multiple times, the total number of possible portfolio combinations is given by a binomial coefficient:
\begin{equation}
M = \binom{n + k - 1}{k} = \frac{(n + k - 1)!}{k! (n - 1)!} \,.
\label{eq:numcombinations}
\end{equation}
For a moderate portfolio size of $n=1000$ and a small investment universe of $k=4$ stocks, the number of possible portfolios is already $M > 10^{10}$. If we extend the number of considered stocks to $k=50$ and keep $n=1000$, the number of possible portfolios grows to $M > 10^{86}$, which is larger than current estimates for the number of atoms in the entire universe.
At the same time, there is no efficient algorithm for finding the optimal combination out of these $M$ combinations on a classical computer, so that only a brute force approach guarantees success. However, most realistic problems are too large to be solved by testing each of the $M$ combinations, so that finding the exact solution of large problems is not feasible on classical machines. Therefore, these problems have been approached using heuristic and approximate methods on classical computers, which do not guarantee an optimal solution~\citep{Mansini1999,Streichert2004,Vielma2008,Li2008,Castro2011}.

In recent years, the rapid progress in manufacturing of quantum processing units (QPUs) and the development of hybrid quantum-classical workflows, not only for universal quantum computers~\citep{Abrams1998,Farhi2014,Zheng2021,Brandhofer2022,Chen2023}, but also for quantum annealers~\citep{Elsokkary2017,Orus2019,Cohen2020,Cohen2020_60Stocks,Philipson2021,Grant2021,Romero2021,Palmer2022,Jacquier2022}, has re-ignited interest in this type of problems. Meanwhile, quantum annealers have been shown to provide quantum advantage for certain classically intractable problems~\citep{King2023} and seem to provide a promising platform for solving quadratic binary optimization and integer quadratic optimization, even in the presence of hard and weak constraints. Based on these prospects, portfolio optimization is a natural application for quantum computing in finance, and in particular quantum annealers. For a broader review of quantum computing applications in finance, see refs.~\onlinecite{Orus2019, Jacquier2022, Herman2023}.

Recently, awareness of environmental, social and governance (ESG) aspects of investing has grown among private and institutional investors alike. A growing number of financial products caters to the growing demand and incorporates ESG aspects into the product design. The trend toward more ESG awareness is likely to get further amplified by regulatory updates on international- and national level. See for example ref.~\onlinecite{Bruno2021} for an overview of ESG regulation in the banking sector across Europe. In January 2023, European authorities agreed on a European implementation of the internationally developed Basel~III update that will result in updated capital requirements regulation (CRR) and capital requirements directive (CRD), including requirements on ESG awareness and inclusion into risk management. Where up to now integrating ESG constraints in investment decisions has been up to individual preferences, it can be expected to become a required standard in the near future in the EU. The inclusion of ESG risk as an additional risk factor besides historical covariance into the Markowitz framework (see eq.~\ref{eq:markowitzutility}) is actively being investigated~\citep{Pedersen2021,Utz2014,Chen2021,Lopez2022}.

Meanwhile, ESG data in different formats are available from a number of data providers such as MSCI, ISS ESG, Refinitiv, Sustainalytics, and others. The approach for establishing ESG ratings varies. Some providers offer ESG ratings or scores that aim to capture investment risks by assessing how effectively a company manages ESG risks to its business (”financial”). Other providers aim to characterize the impact of a corporation on the environmental, social and governance dimensions, with the goal of facilitating informed decisions for investors (“impact”). The ISS ESG data used in this analysis captures both the financial and impact aspects of ESG ratings. The scores can be given on the level of environmental, social and governance dimensions, with focus on smaller sub-areas, or as a single aggregate score on company level, which seeks to represent the average of all relevant aspects.  For a critical review of available data sets and methodologies see refs.~\onlinecite{Larcker2022,Berg2022}.

How ESG ratings should be best included into the Markowitz framework is an open question. Both inclusion of the expected ESG score into the vector of expected returns~\citep{Alessandrini2021,Shushi2022,Varmaz2022,Lauria2022} and optimizing the ESG score in the form of a multi-objective optimization~\citep{Utz2014,Hirschberger2013,Cesarone2022,DeSpiegeleer2023} have been investigated in the literature. Including the ESG score into the returns vector is intrinsically ambiguous, since it compares ESG score and monetary returns as if these quantities had the same units. This introduces a conversion law between returns and ESG scores, which depends on the exact form of the ESG score data, which may differ between various providers. However, it would be preferable to have a unique framework for incorporating ESG scores into the Markowitz utility function (see eq.~\ref{eq:markowitzutility}). The multi-objective optimization approach, on the other hand, may be unable to control the interplay between returns, variance and ESG performance, depending on the exact implementation.

In this work, we extend the Markowitz portfolio theory to include the ESG scores directly in the utility function in a way that avoids the ambiguity in relation to the returns. Furthermore, we can investigate and control the interplay between returns, variance and ESG performance. Our formulation is applicable to standard (continuous) and also discrete mean-variance (Markowitz) portfolio optimization, to allow for application to realistic scenarios. We demonstrate the feasibility of our method on classical computers for the continuous case and quantum annealers for the discrete portfolio optimization case. Results are based on real market data of selected stocks from EURO STOXX 50 index as well as actual respective ESG scores from ISS ESG.

The paper is divided as follows:
Section~\ref{sec:results} contains the main results of our study.
In subsection~\ref{sec:MPTtoDMPT} we establish the correct normalization approach for the discrete Markowitz problem, so that solutions for the continuous and the discrete formulation may be compared. We provide a relationship between the total number of stocks in the portfolio and the risk-aversion parameter, which needs to be considered.
In subsection~\ref{sec:DMPTBudget} we introduce a budget constraint into the discrete portfolio problem, so that realistic scenarios with limited budget may be investigated. We compare the usual rounding approach to direct search of discrete optimal portfolios and find that rounding produces sub-optimal portfolios for small to medium investment budgets.
In subsection~\ref{sec:ESGutilityFunction} we introduce a novel framework for including ESG scores into both continuous and discrete Markowitz portfolio optimization, which is applicable even to ESG data with heterogeneous scales.
In section~\ref{sec:discussion} we discuss our results and potential implications for ESG-aware investors.
Finally, in section~\ref{sec:conclusions} we summarize our results and provide an outlook on future research topics.

\section{Results}
\label{sec:results}
\subsection{Discrete Markowitz portfolio theory and role of the risk-aversion parameter}
\label{sec:MPTtoDMPT}
Here we investigate the connection between the continuous and the discrete Markowitz portfolio theory. Naively, one could expect that the discrete approach should yield the same results as the continuous version for large portfolios, where the discreteness becomes less relevant. As mentioned in the introduction, already the single period DMPT is an NP-hard problem~\citep{Castro2011, Rosenberg2015}. 

To explore this connection in detail, we formalize the DMPT problem with a fixed number of stocks in the portfolio in the following way:
\begin{equation}
\label{eq:dmptutility}
\min \left\{ Q_d(\vec{r}, \Sigma, \phi,N_\text{tot}) \right\} = \min \left\{ \frac{\phi}{2} \vec{x}^T \Sigma \vec{x} - \vec{r}^T \vec{x} \right\} \,.
\end{equation}
The crucial difference to the continuous case (see eq.~\ref{eq:markowitzutility}) lies in the discrete nature of the constraints:
\begin{equation}
\begin{split}
\label{eq:dmptutilityconstraints}
&\vec{x}=(x_1, \dots, x_k) \quad \text{with} \quad x_i \in \mathbb{N} \quad \forall i,\\
&\sum_{i=1}^k x_i  = N_\text{tot} \,.
\end{split}
\end{equation}
Note that the return vector $\vec{r}$ and the covariance matrix $\Sigma$ do not change their meaning, since these quantities are dimensionless. Therefore, no special care has to be taken when interpreting return vector $\vec{r}$ or covariance matrix $\Sigma$ in the continuous vs.~the discrete case.

If the raw solution of this naive approach (eqs.~\ref{eq:dmptutility} and \ref{eq:dmptutilityconstraints}) is denoted as $\vec{X}_\text{naive}$, we can calculate the relative portfolio weights $\vec{x}_{d, \text{naive}}$, which may be compared to the solution $\vec{x}_c$ from the continuous case, by dividing through the portfolio size $N_\text{tot}$:
\begin{equation}
\vec{x}_{d, \text{naive}} = \frac{\vec{X}_\text{naive}}{N_\text{tot}} \,.
\label{eq:dmptnaiveweights}
\end{equation}
If we calculate the Euclidean distance between the continuous solution $\vec{x}_c$ and these naive weights from eq.~\ref{eq:dmptnaiveweights}, we would expect the difference from $\vec{x}_c$ to vanish with increasing $N_\text{tot}$:
\begin{equation}
\lim_{N_\text{tot} \to \infty} || \vec{x}_c - \vec{x}_{d, \text{naive}} ||_2 = 0 \,.
\label{eq:dmptnaivedifflimit}
\end{equation}
We calculated this difference with the formalism described so far. The continuous solution was extracted using the \texttt{CVXPY}~\citep{Diamond2016, Agrawal2018} software package for the Python programming language. For solving the discrete portfolio optimization problem, one could use a heuristic classical algorithm~\citep{Garcia2022}, an algorithm for gate-based quantum computers~\citep{Mugel2022, Shunza2023}, a quantum-inspired approximate algorithm for classical computers~\citep{Mugel2022} or a quantum annealer~\citep{Sakuler2023, Jacquier2022}. For an overview of the use of various computing approaches in portfolio optimization, see ref.~\onlinecite{Buonaiuto2023}.

D-Wave quantum annealers implement the Ising model, known from theoretical physics, in a specialized quantum processing unit. These annealers are not universal quantum computers. Therefore, their applicability is limited to problems, which can be represented in terms of the Ising model~\citep{Lucas2014}. The solution of the optimization problem is extracted via a physical annealing process, which gradually cools the quantum processing unit down to temperatures close to absolute zero. Subsequently, the quantum state of the system is measured and translated back to the original problem space.

We have decided to use a quantum annealer, because the discrete portfolio optimization problem can be rewritten as an Ising model.~\citep{Lucas2014, Jacquier2022} Therefore, the quantum annealer is a natural choice when solving discrete portfolio problems. Other previously mentioned approaches are also viable~\citep{Buonaiuto2023}, but either cannot reach the problem sizes considered here or have no guarantee of providing an optimal solution. Nevertheless, heuristic approaches may yield very good results as demonstrated in ref.~\onlinecite{Garcia2022}.

The Ising model is formulated in terms of discrete variables, which represent magnetic moments. These can be in one of two quantum states $s_i \in \{+1, -1\}$. A simple transformation allows us to convert these magnetic moments into zeros and ones $a_i \in \{0, 1 \}$, which can be used to represent integer numbers in binary encoding:
\begin{equation}
a_i = \frac{s_i + 1}{2}
\label{eq:isingtoqubo}
\end{equation}

This transformation between integer optimization problems and the Ising model is well-known~\citep{Lucas2014} and automatically carried out in various software packages like D-Wave \texttt{Ocean}.~\footnote{\url{https://ocean.dwavesys.com/} (accessed on 21 March 2024)} The optimization problem can be entered into \texttt{Ocean} in a declarative way using a domain-specific language. In particular, this means that no imperatively formulated solution algorithm is required. This software package also handles the transformation of constraints into penalty terms in a proprietary way. For details on the technical implementation of D-Wave solvers, see ref.~\onlinecite{DwaveHybridSolver2021}. Note, however, that going beyond long-only portfolios would require a type of optimization constraint, which is currently not supported by D-Wave software packages.

Now we estimate a theoretical upper bound to the number of qubits required by our approach. Since D-Wave \texttt{Ocean} uses binary encoding for integer variables, the upper bound for the required number of qubits $N_\text{qubit}$ scales with the logarithm of the portfolio size $N_\text{tot}$ and linearly in the number of assets $k$ within our investment universe: 
\begin{equation}
N_\text{qubit} \leq k \cdot \Big( \log_2 \left( N_\text{tot} + 1 \right) + 1 \Big) \,.
\label{eq:nqubits}
\end{equation}
Of course, the proprietary algorithm of D-Wave may require an overhead of additional qubits to encode the problem on real-world hardware. Unfortunately, these details are not public and cannot be investigated here further.

\begin{figure}[tb]
\includegraphics[width=\linewidth]{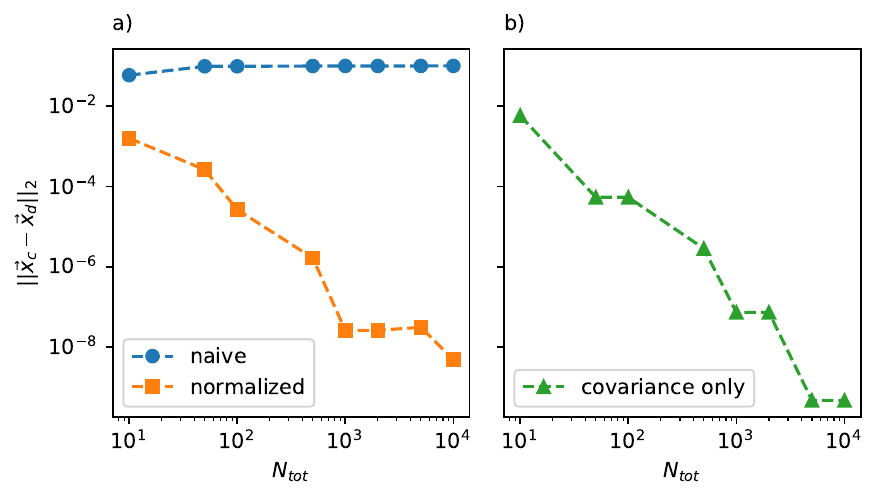}
\caption{Euclidean norm of the difference vector between optimal relative portfolio weights $\vec{x}$ for the continuous ($\vec{x}_c$) and discrete optimization case ($\vec{x}_d$). The risk-aversion parameter is set to $\phi=8$, but different choices of $\phi > 0$ give similar results. The investment universe comprises BMW (ISIN DE0005190003), Deutsche Post (ISIN DE0005552004), Deutsche Telekom (ISIN DE0005557508) and Infineon (ISIN DE0006231004). Data is taken from the period between 1 January 2010 and 1 January 2021. Lines are guides to the eye. a) Difference between continuous solution and naive discrete approach (circles) as well as the difference between continuous and normalized discrete solution (squares). Obviously, the naive approach does not converge to the continuous solution, even for very large portfolios. The normalized discrete approach converges to the well-known continuous solution for large portfolios. The remaining differences in portfolio composition are purely due to the discreteness. b) Difference between continuous and naive discrete solutions for the modified utility function $Q_\text{mod} = \vec{x}^T \Sigma \vec{x}$, which only includes the covariance term. It is clearly visible that both the continuous and discrete approaches converge to the same minimum variance portfolio for this modified utility function $Q_\text{mod}$. Also here, the remaining differences in portfolio composition are purely due to the discreteness. }
\label{fig:utilityvariations}
\end{figure}

The results of our calculations using \texttt{CVXPY} for the continuous problem and D-Wave \texttt{Ocean} for the discrete problem are shown in Fig.~\ref{fig:utilityvariations}a). We realized that the difference in eq.~\ref{eq:dmptnaivedifflimit} does not converge to zero with growing portfolio size. That is the case, because risk-aversion parameters for the continuous and discrete portfolio cases are not directly comparable. This phenomenon does not depend on the exact value of $\phi > 0$.

If we view the continuous problem of eq.~\ref{eq:markowitzutility} as a particular discrete problem, in which the solution vector $\vec{x}$ is rescaled by $1 / N_\text{tot}$ and the limit $N_\text{tot} \to \infty$ is applied, we can write it in the following way:
\begin{equation}
\label{eq:viewcontinuousasdiscretelimit}
\begin{array}{rl}
& \phantom{=} \min \left\{Q_c(\vec{r}, \Sigma,\phi) \right\}\\[12pt]
&= \min \left\{ \lim\limits_{N_\text{tot} \to \infty} \left(\frac{\phi}{2} \frac{\vec{x}^T \Sigma \vec{x}}{N_\text{tot}^2} - \frac{\vec{r}^T \vec x}{N_\text{tot}} \right) \right\} \\[12pt]
&= \min \left\{ \lim\limits_{N_\text{tot} \to \infty} \left(\frac{1}{N_\text{tot}} \right) \left( (\frac{1}{N_\text{tot}}) \frac{\phi}{2} \vec{x}^T \Sigma \vec{x} - \vec{r}^T \vec{x} \right) \right\} \,.
\end{array}
\end{equation}
The constraints are the same as in eq.~\ref{eq:dmptutilityconstraints}. It is clear that the discrete problem of eq.~\ref{eq:viewcontinuousasdiscretelimit} can only converge to the continuous problem of eq.~\ref{eq:markowitzutility} if the additional factor of $1 / N_\text{tot}$ in front of the covariance term is absorbed into the risk-aversion parameter. Hence, the risk-aversion parameter of the discrete case $\phi_d$ is connected to the risk-aversion parameter of the continuous case $\phi_c$ in the following way:
\begin{equation}
\label{eq:riskaversionrenormalize}
    \phi_{d} = \frac{\phi_c}{N_\text{tot}} \,.
\end{equation}
Therefore, we need to respect the mapping for the normalized risk-aversion parameter $\phi$ in eq.~\ref{eq:riskaversionrenormalize} if we want to compare portfolios from continuous and discrete optimization. Doing this correctly and re-calculating the difference in eq.~\ref{eq:dmptnaivedifflimit} with the normalized risk-aversion parameter, one obtains the second curve in Fig.~\ref{fig:utilityvariations}a), which clearly converges to zero for a large number of stocks $N_\text{tot}$.

\begin{figure}[tb]
\includegraphics[width=\linewidth]{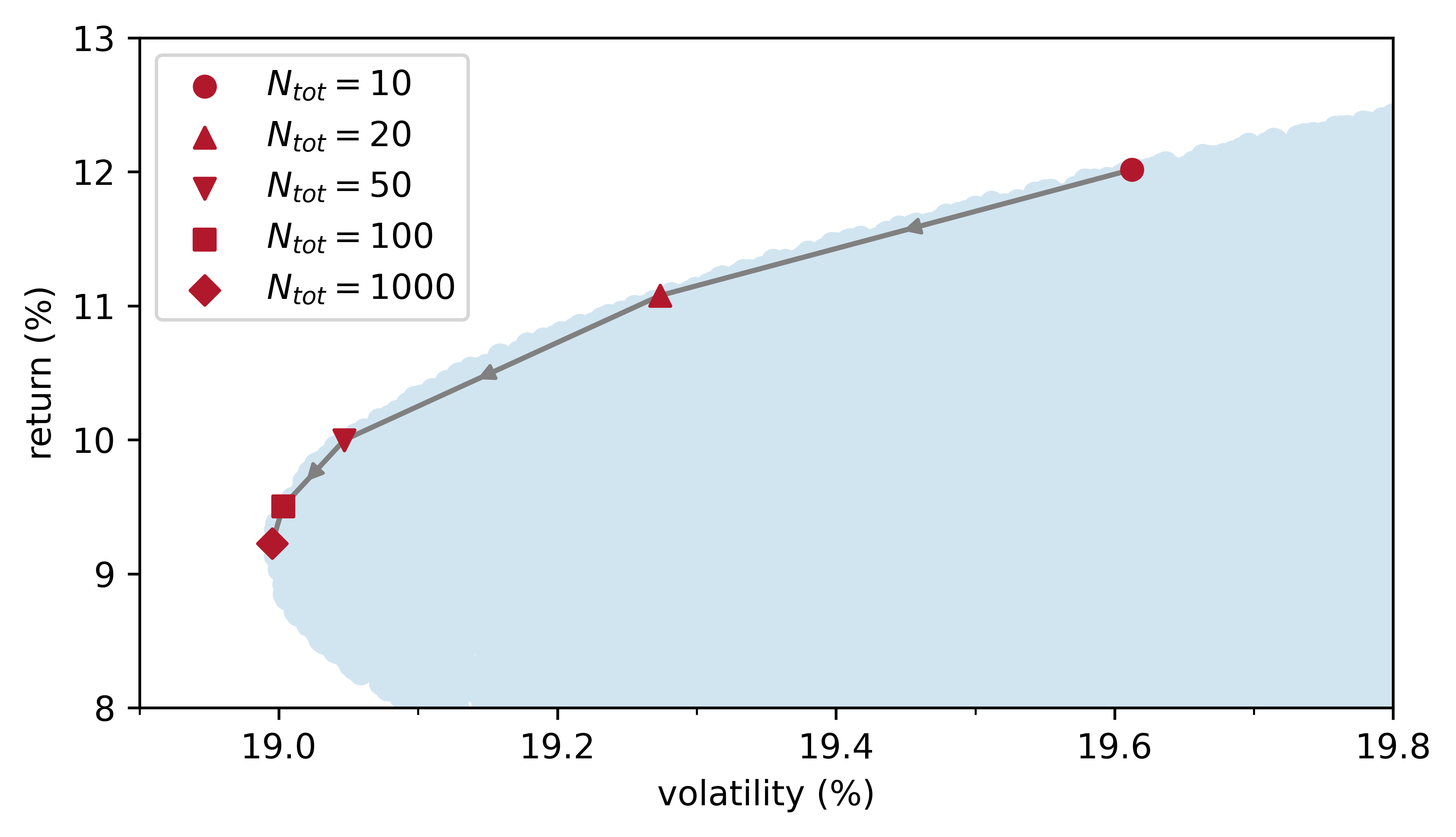}
\caption{Portfolio positions in volatility-return space for the naive discrete optimization approach as a function of the total number of stocks $N_\text{tot}$ in the portfolio. The risk-aversion parameter is set to $\phi=1$. The investment universe comprises BMW (ISIN DE0005190003), Deutsche Post (ISIN DE0005552004), Deutsche Telekom (ISIN DE0005557508) and Infineon (ISIN DE0006231004). The light blue background is generated by randomly sampling the space of possible portfolios. The upper boundary of the light blue area is commonly called 'efficient frontier'. Data is taken from the period between 1 January 2010 and 1 January 2021. Lines and arrows are guides to the eye.}
\label{fig:frontier_norescale_discrete}
\end{figure}

It is also instructive to examine the solutions of the naive approach for different portfolio sizes (without renormalizing $\phi$) and their position in volatility-return space. This is shown in Fig.~\ref{fig:frontier_norescale_discrete}. All solutions lie on the 'efficient frontier', as the surface of maximum return as a function of volatility is commonly called. This efficient frontier in the background was generated by sampling random portfolio compositions. 

As we can see, the solutions from the naive discrete approach trend towards the minimum-variance solution if we naively fix $\phi = 1$. The reason is clear from eq.~\ref{eq:riskaversionrenormalize}: with $N_\text{tot} \gg 1$, we should have adapted the risk-aversion parameter to the portfolio size. For example we should have used $\phi = 1 / 1000$ for $N_\text{tot} = 1000$ in order to obtain comparable solutions. Fixing $\phi = 1$ irrespective of the portfolio size leads to portfolios, for which risk-aversion becomes increasingly important as the size of the portfolio grows. Therefore, the naive approach always converges to the minimum-variance portfolio for $N_\text{tot} \gg 1$. Also note how the scale of variations in volatility in Fig.~\ref{fig:frontier_norescale_discrete} is small, already for $N_\text{tot} = 10$, due to the over-emphasis on risk-aversion. The convergence to the minimum variance portfolio happens rapidly as a function of the number of stocks $N_\text{tot}$. Already at $N_\text{tot} = 1000$ the composition is practically indistinguishable from the minimum variance portfolio.

As a final test, we have carried out another calculation, in which we have neglected the term related to maximizing the return in the utility function (see eq.~\ref{eq:dmptutility}). If only the covariance term is considered, the optimization should always yield the minimum-variance portfolio irrespective of the portfolio size. This is clearly the case, as shown in Fig.~\ref{fig:utilityvariations}b). The remaining difference in portfolio compositions between continuous and discrete solutions is purely due to the discrete stock allocation in the latter case. Thus, we have shown that renormalizing the risk-aversion parameter $\phi$ according to eq.~\ref{eq:riskaversionrenormalize} is crucial.

Now that we have established a discrete portfolio optimization approach, which is comparable to the well-known continuous approach, we introduce budget constraints in the following subsection to mimic realistic portfolio selection problems.

\subsection{Discrete Markowitz portfolio theory with limited investment budget}
\label{sec:DMPTBudget}
So far, we have only solved the portfolio problem with a limited number of stocks. In practice, the number of stocks that can be purchased is usually not limited directly, but indirectly via the total available investment budget. To make our study more realistic, we now fix the total available investment budget. This means that the algorithm will not optimize different stocks like for like, but rather optimize portfolios with many low-price stocks versus portfolios with few high-price stocks.

As explained in section~\ref{sec:MPTtoDMPT}, the total number of stocks in discrete Markowitz portfolio theory plays a crucial role in the risk-aversion parameter, which determines the compromise between risk and return of the portfolio. With the risk-aversion parameter for the continuous portfolio $\phi_c$, we write the utility function for the discrete portfolio theory in the following way:
\begin{equation}
\label{eq:budget_utility}
Q_d(\vec{r}, \Sigma, \phi_c,N_\text{tot}) = \frac{\phi_c}{2 N_\text{tot}} \vec{x}^T \Sigma \vec{x} - \vec{r}^T \vec{x} \,.
\end{equation}
The minimization of this utility function is subject to the following constraints:
\begin{equation}
\begin{split}
\label{eq:utilitybudgetconstraints}
&\vec{x}=(x_1, \dots, x_k) \quad \text{with} \quad x_i \in \mathbb{N} \quad \forall i,\\
&\sum_{i=1}^k x_i  = N_\text{tot},\\
& \vec{p}^T \vec{x} - B \leq 0 \,.
\end{split}
\end{equation}
Here, $\vec{p}$ is the vector, which contains the price per stock for each stock. Therefore, $\vec{p}^T \vec{x}$ is the initial value of the portfolio. $B$ is the initially available investment budget. In this sense, we constrain the optimization to the space of those portfolios, which can be purchased with the initially available budget. Since we also maximize return via the utility function (eq.~\ref{eq:budget_utility}), the algorithm will yield portfolios, which use the available budget to the maximum extent.

In practice, we will study the problem defined by eqs.~\ref{eq:budget_utility} and \ref{eq:utilitybudgetconstraints} at a fixed number of stocks $N_\text{tot}$. If the number of stocks $N_\text{tot}$ is chosen too small, the initial portfolio value will be far below the initial budget $B$. If we choose a too large number $N_\text{tot}$, the number of possible portfolio combinations will exceed the capabilities of contemporary quantum hardware. Therefore, we start with low $N_\text{tot}$ and gradually increase this number until the difference between portfolio value and available budget $\vec{p}^T \vec{x} - B$ becomes sufficiently small. As we will see, this approach yields good results even in realistic settings.

Of course, we would like to compare these discrete solutions to portfolios that are based on the usual continuous Markowitz theory. In the continuous case, the solution $\vec{x}_c$ provides a relative allocation of the available investment budget to the respective stocks. The actual portfolio is then usually constructed by multiplying the relative weights $\vec{x}_c$ with the available budget $B$. This gives the budget, which is allocated to each stock. To obtain the integer number of stocks that has to be bought for each sort, one divides by the price of the respective stock and rounds to the next integer, which is denoted as $\lfloor \rceil$. Therefore, we can write the integer portfolio composition based on the rounding approach as:
\begin{equation}
\label{eq:discretefromrounding}
\left( \vec{x}_{c, r} \right)_i = \left\lfloor \frac{B \cdot \left( \vec{x}_{c}\right)_i}{p_i} \right\rceil \,.
\end{equation}
Here, $(\vec{x}_{c})_i$ is the i-th component of the vector of relative allocation from the continuous Markowitz theory and $p_i$ is the price of the i-th stock. Interestingly, the rounding approach according to eq.~\ref{eq:discretefromrounding} yields portfolio compositions, which are substantially different from the discrete approach using eqs.~\ref{eq:budget_utility} and \ref{eq:utilitybudgetconstraints}, even if consistent values for the risk-aversion parameter $\phi_c$ are used. Remember that these two approaches only coincide in the limit of infinite available budget, as explained in subsection~\ref{sec:MPTtoDMPT}.

\begin{figure}[tb]
\includegraphics[width=\linewidth]{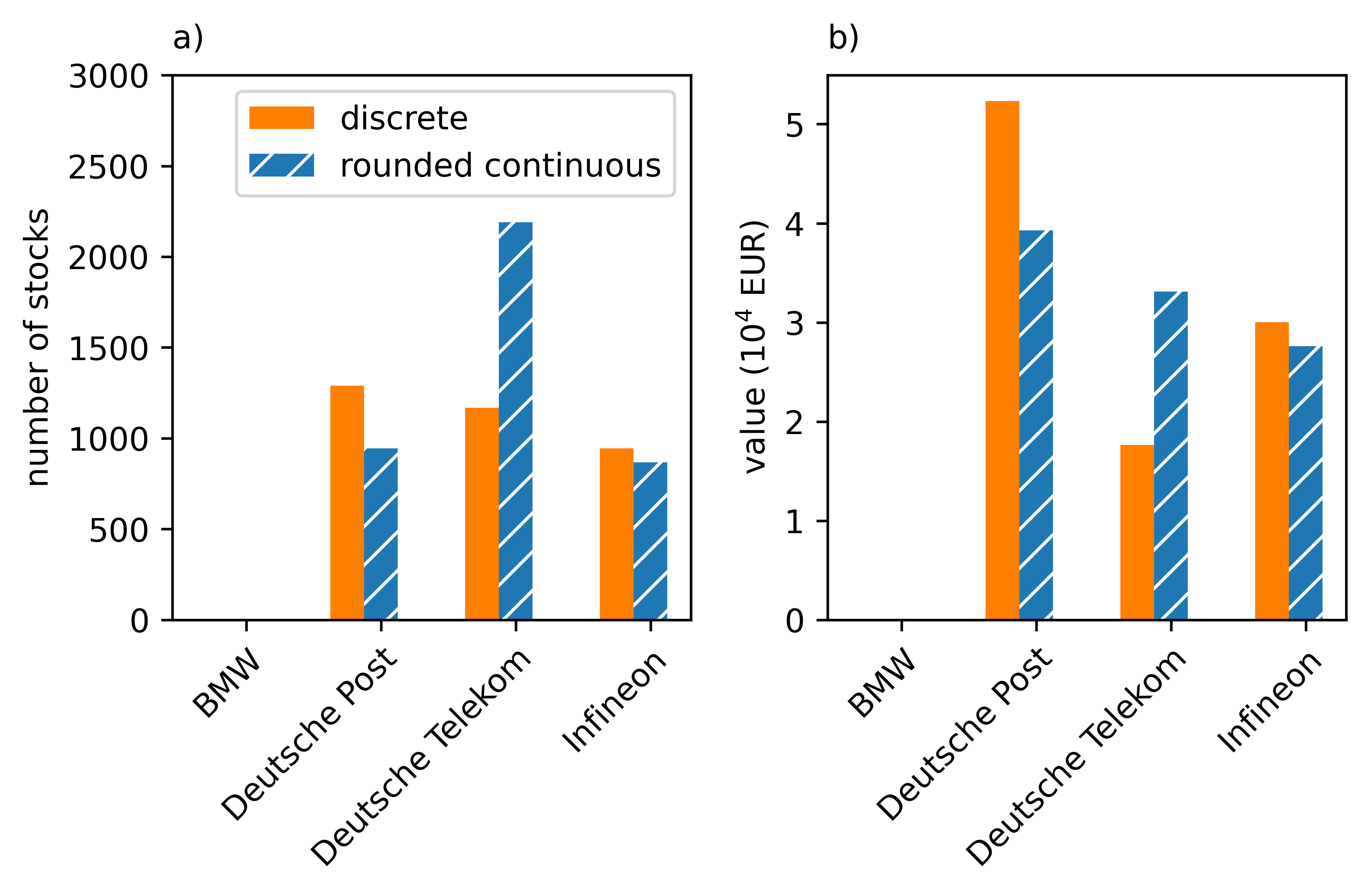}
\caption{Best portfolio compositions for a budget of $B = 100000$~€ and risk-aversion parameter of $\phi=8$. The discrete solution is obtained by minimizing the utility function in eq.~\ref{eq:budget_utility} under the constraints of eq.~\ref{eq:utilitybudgetconstraints}. We use $N_\text{tot} = 3401$. The continuous results were obtained by multiplying the relative allocation by the available budget and rounded to integer stocks via eq.~\ref{eq:discretefromrounding}, which results in $N_\text{tot}=4026$. The investment universe comprises BMW (ISIN DE0005190003), Deutsche Post (ISIN DE0005552004), Deutsche Telekom (ISIN DE0005557508) and Infineon (ISIN DE0006231004). Data is taken from the period between 1 January 2010 and 1 January 2021. a) Portfolio composition in terms of number of stocks. b) Portfolio composition in terms of Euro value.}
\label{fig:budgetcomposition}
\end{figure}

We have carried out continuous and discrete portfolio optimization with a risk-aversion parameter of $\phi_c = 8$ and a total investment budget of $B = 100000$~€. For simplicity, the investment universe is again limited to BMW (ISIN DE0005190003), Deutsche Post (ISIN DE0005552004), Deutsche Telekom (ISIN DE0005557508) and Infineon (ISIN DE0006231004). In the discrete case we have used $N_\text{tot}=3401$, which produces an initial portfolio value of $\vec{p}^T \vec{x}_d = 99999.87$~€ for the optimal solution. The rounding approach (see eq.~\ref{eq:discretefromrounding}) may of course slightly violate the budget constraint. Thus, the rounded solution yields an initial portfolio value of $\vec{p}^T \vec{x}_{c, r} = 100006.32$~€ and a total number of stocks $N_\text{tot}=4026$. The larger number of stocks for the rounded continuous case appears, because the standard Markowitz approach puts a large relative weight on Deutsche Telekom, which has the lowest Euro value per stock within the considered investment universe. This means that a larger number of these stocks will be bought with the available budget.

The resulting portfolio compositions for the discrete and rounded continuous cases are shown in Fig.~\ref{fig:budgetcomposition}, both in terms of the number of stocks bought per ISIN and the invested budget per ISIN. We observe that the respective portfolio compositions are strikingly different. The rounded continuous approach yields a solution, which is well diversified in terms of allocated budget. The discrete approach on the other hand yields a portfolio, which is slightly more concentrated in terms of budget allocation. This effect is likely due to the strict budget constraint in the discrete case, which forces the optimization to pick allocations that fit the specific budget constraints.

\begin{figure}[tb]
\includegraphics[width=\linewidth]{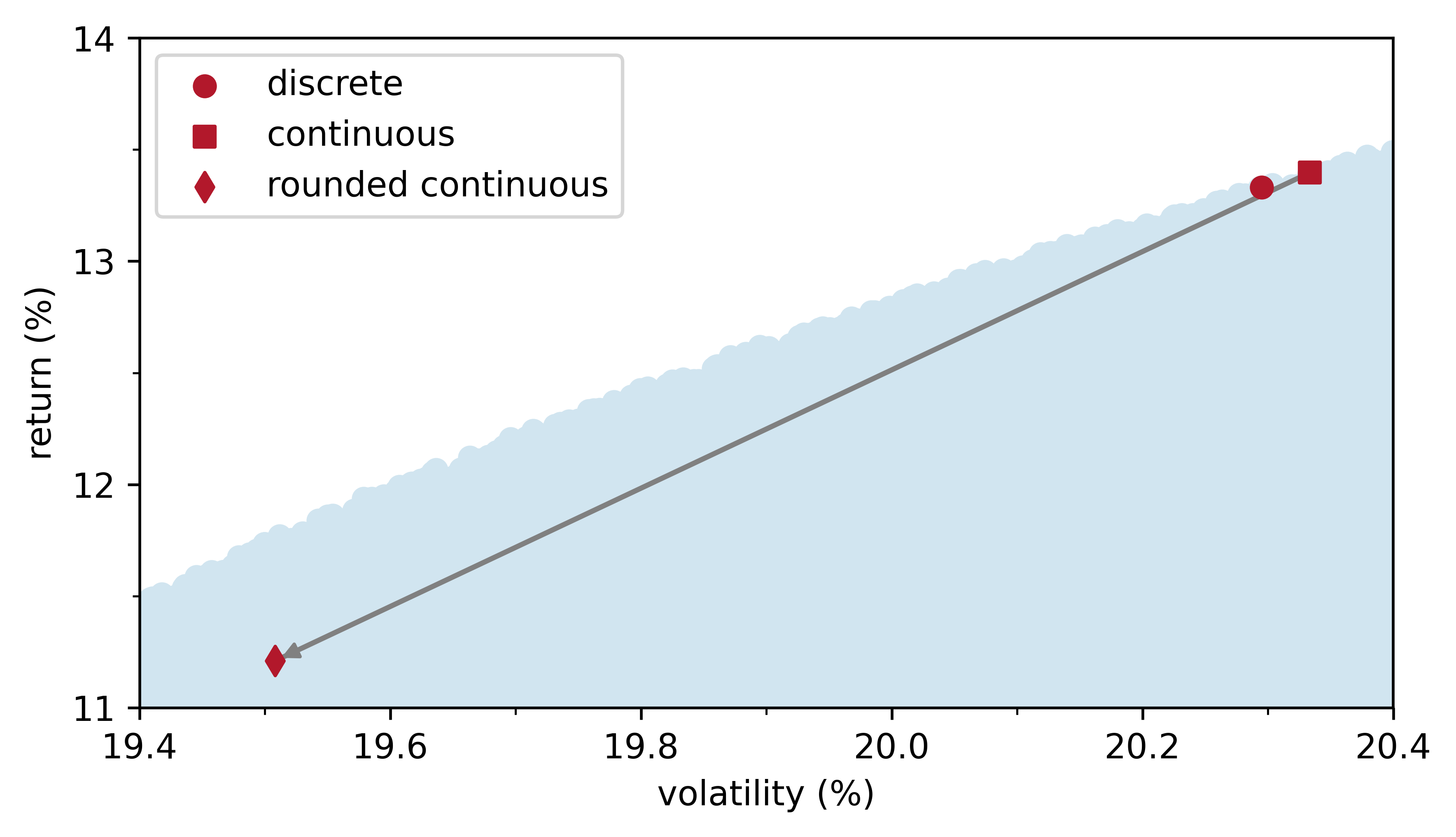}
\caption{Position of the best portfolio compositions in volatility-return space for a budget of $B = 100000$~€ and risk-aversion parameter of $\phi=8$. The discrete solution is obtained by minimizing the utility function in eq.~\ref{eq:budget_utility} under the constraints of eq.~\ref{eq:utilitybudgetconstraints}. We use $N_\text{tot} = 3401$. The continuous results were obtained by multiplying the relative allocation by the available budget and rounded to integer stocks via eq.~\ref{eq:discretefromrounding}, which results in $N_\text{tot}=4026$. The investment universe comprises BMW (ISIN DE0005190003), Deutsche Post (ISIN DE0005552004), Deutsche Telekom (ISIN DE0005557508) and Infineon (ISIN DE0006231004). Data is taken from the period between 1 January 2010 and 1 January 2021. The discrete solution is clearly at the efficient frontier, while the rounded continuous solution is visibly sub-optimal. Arrows are guides to the eye.}
\label{fig:budgetfrontier}
\end{figure}

We also investigated the position of the obtained portfolios in volatility-return space (see Fig.~\ref{fig:budgetfrontier}). The discrete solution is right at the efficient frontier, i.e. it yields an optimal return for the given volatility. The rounded continuous portfolio has a lower volatility, but also yields a significantly lower than optimal return. The deviation of nearly 2 percentage points in return is larger than one may expect from the seemingly harmless rounding approach. The effects of the rounding observed here are likely relevant in practical applications. In fact, one can expect even larger deviations for portfolios with a larger number of components.

Our results clearly show that the continuous and discrete approaches only converge to identical results in the limit of infinite portfolio size. For a limited investment budget, the approach of minimizing the utility function for the discrete case directly on the quantum computer yields results, which are far superior to the widely used rounding method based on the standard Markowitz approach, even for moderately sized portfolios and limited investment universe. We expect that the quantum computing approach will have even stronger appeal for large investment universes, since the discreteness of individual components will play an even more important role there.

Now that we have established the superiority of the quantum computing approach in the case of limited budget, we come to the main idea of our study: the inclusion of ESG data into the discrete portfolio optimization problem.

\subsection{Incorporation of ESG data into Markowitz portfolio theory}
\label{sec:ESGutilityFunction}
We have to address two questions in order to include ESG data into Markowitz portfolio optimization: i) how to classify portfolios in terms of ESG scores and ii) how to incorporate such information into the optimization scheme. The current literature on this topic can be divided into two main approaches: the most commonly found way of including ESG data is to constrain the Markowitz utility function, so that it yields a portfolio with the weighted average of the expected ESG scores~\citep{Maree2022,Chen2021,Branda2015,Lopez2022,Utz2014,Varmaz2022,Alessandrini2021,DeSpiegeleer2023,Cesarone2022,Hirschberger2013,Shushi2022}, which actually constrains the possible portfolio compositions. The second approach~\citep{Lauria2022} employs an affine transformation between returns and ESG scores, which is controlled by an additional parameter.

Obviously, the composition-weighted average of expected ESG scores is not the only property that can be used to classify portfolios. In this subsection we introduce a novel scheme for classifying portfolios in terms of ESG score, which we incorporate into the discrete portfolio optimization scheme explained in subsection~\ref{sec:DMPTBudget}.

We assume that the value of the ESG score $S$ in every scoring system is bounded by the best and worst possible scores $S \in [S^-, S^+]$. Let us consider a relative portfolio composition $\vec{\pi}$ with respect to the ESG scores of a given scoring system. Since the entries of $\vec{\pi}$ are non-negative and their sum is one, the entries of $\vec{\pi}$ can be interpreted as a probability distribution. As a reference point we take a portfolio, which only contains stocks with the best possible ESG score $S^+$. With the result of Wasserstein~\citep{Peyre2019}, distances of other relative portfolio compositions with respect to this best possible portfolio can be calculated. Note that there may be multiple portfolio compositions in terms of stock allocation, which possess the best possible score $S^+$, e.g. if more than one stock in the investment universe has the best possible score. However, this possible degeneracy is irrelevant in our approach, as we will see.

If we use the Wasserstein metric in the case of two one-dimensional sets of measurements and take the limit of infinite number of observations~\citep{Villani2021,Kolouri2017}, we can write the Wasserstein $p$-distance between a given relative portfolio composition $\vec{\pi}$ and the best possible portfolio in the following way:
\begin{equation}
\label{eq:wasserstein}
D_\text{ESG}(p, \vec{\pi})=\left[ \sum_i \left( \pi_i \cdot \left| S^+- \tilde{S}_i \right|^p \right) \right]^{1/p} \,. 
\end{equation}
Here, $i$ enumerates the possible values $\tilde{S}_i$ of the ESG score within the given portfolio $\vec{\pi}$. This vector contains the relative number of stock allocations to the respective ESG score $\tilde{S}_i$. $\pi_i$ is the i-th component of the vector $\vec{\pi}$. $p\in[1,+\infty)$ is the parameter of the Wasserstein $p$-distance. Note how the exact composition in terms of stocks is irrelevant in this approach. The distance measure $D_\text{ESG}$ is only sensitive to the ESG scores of the respective constituents. Therefore, different allocations of stocks with the same ESG score do not affect $D_\text{ESG}$. Also note that comparison to a best possible individual allocation would have required us to know this specific portfolio. This target portfolio is, however, in general unknown. The point of our method is to find it. Hence, we have chosen an approach in which knowledge of this hard to find solution is not required.

If all constituents of a portfolio $\vec{\pi}$ have the best possible score $S^+$, the distance measure is $D_\text{ESG}(p, \vec{\pi}) = 0$. If all constituents of a portfolio $\vec{\pi}$ have the worst possible score $S^-$, the distance measure is $D_\text{ESG}(p, \vec{\pi}) = \left| S^+- S^- \right|$. Therefore, all other portfolios have $D_\text{ESG}(p) \in \left[ 0,  \left| S^+- S^- \right| \right]$ independently of $p$. In particular, if for two given portfolios $\vec{\pi}_1$ and $\vec{\pi}_2$ we have $D_\text{ESG}(p, \vec{\pi}_1) < D_\text{ESG}(p, \vec{\pi}_2)$, then $\vec{\pi}_1$ has a better ESG score.

For $p=1$ our result in eq.~\ref{eq:wasserstein} becomes the weighted average (up to a constant factor). Therefore, we may view eq.~\ref{eq:wasserstein} as a generalized framework for classifying portfolios in terms of ESG scores. This framework does not depend on whether the best score $S^+$ has the lowest or the highest value in the respective scoring system. Also note that our distance measure may be generalized to work with heterogeneous data from different ESG data providers by using the relative distance of the single portfolio component within its pertinent ESG score range, by extending eq.~\ref{eq:wasserstein} with an additional normalization factor:
\begin{equation}
    D_\text{ESG}(p, \vec{\pi})=\left[ \sum_i \left( \pi_i \cdot \frac{\left| S^+(\pi_i) - \tilde{S}_i(\pi_i) \right|^p}{\left| S^+(\pi_i) - S^-(\pi_i) \right|^p} \right) \right]^{1/p} \, 
\end{equation}
Here $S^+(\pi_i),~S^-(\pi),~\tilde{S_i}(\pi_i)$ indicate respectively the best, the lowest and the spot score in the ESG system pertinent to the component $\pi_i$. Although our methodology would enable us to mix multiple ESG scoring systems, we do not expand upon this topic in the present manuscript and leave it for future research instead. In the present manuscript we only use ESG data provided by ISS ESG.

\begin{figure}[t]
\includegraphics[width=\linewidth]{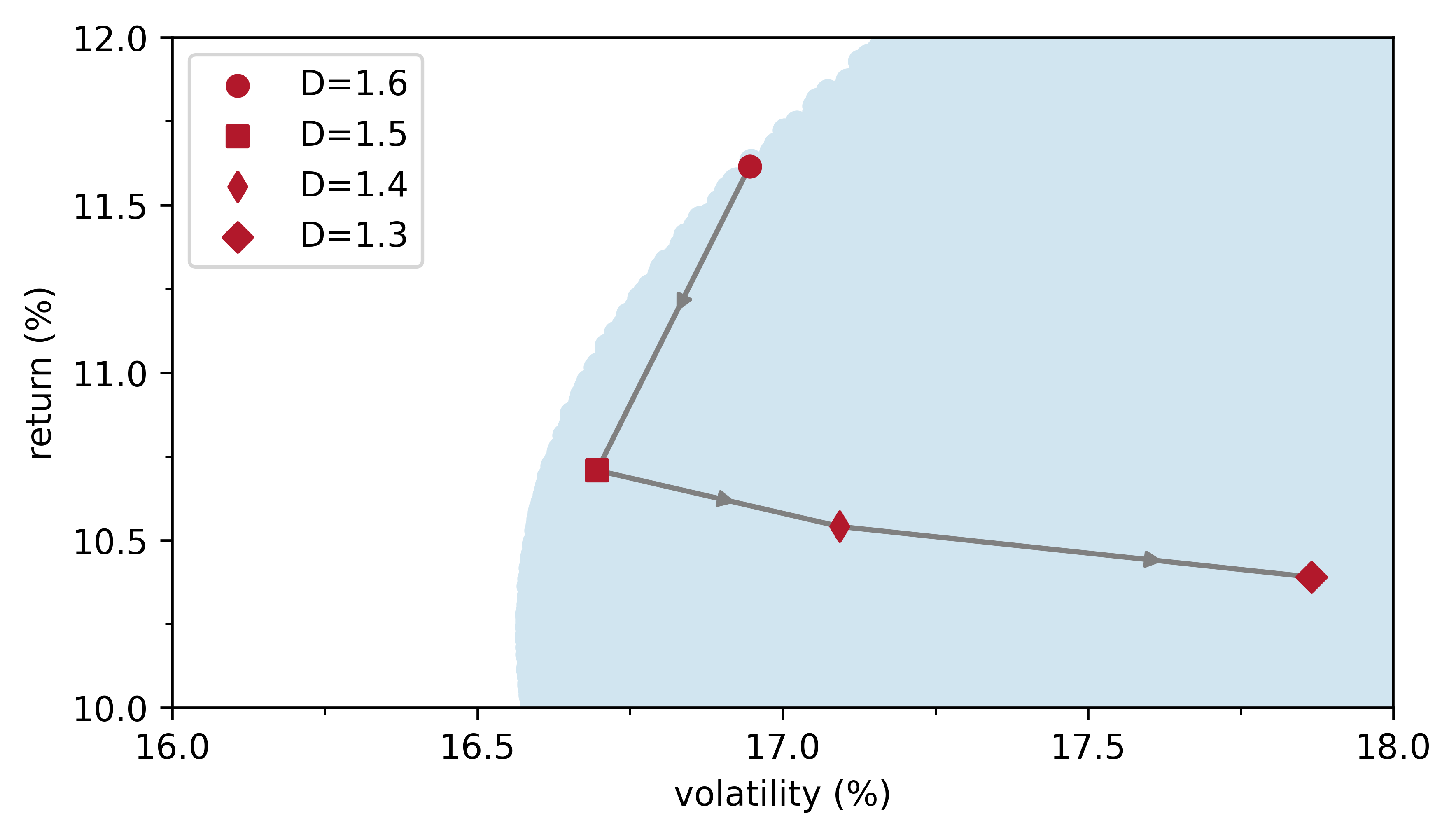}
\caption{Position of the best discrete portfolio compositions in volatility-return space for a budget of $B = 100000$~€, a risk-aversion parameter of $\phi=8$ and different values of maximum allowed ESG distance $D$. The ESG data were provided by ISS ESG. The grading system is on a scale from $4$ to $1$, where a higher number indicates better ESG performance. The investment universe comprises Deutsche Telekom (ISIN DE0005557508), SAP (ISIN DE0007164600), Intesa Sanpaolo (ISIN IT0000072618) and EssilorLuxottica (ISIN FR0000121667). Those are given in order from highest to lowest ESG score. Data is taken from the period between 1 January 2010 and 1 January 2021. The solution portfolios move away from the efficient frontier as we restrict them into a space that becomes gradually tighter around the best possible ESG score. Arrows are guides to the eye.}
\label{fig:positionESGfrontier}
\end{figure}

So far we have written the distance measure in terms of the relative composition with respect to the ESG score. In order to include the ESG data into the discrete optimization framework, we need to establish the ESG distance measure in terms of the composition with respect to the allocation of individual stocks. It is easy to show that eq.~\ref{eq:wasserstein} can equivalently be written as:
\begin{equation}
\label{eq:wassersteinstocks}
D_\text{ESG}(p, \vec{x})=\left[ \sum_{i=1}^k \left( \frac{x_i}{N_\text{tot}} \cdot \left| S^+- S_i \right|^p \right) \right]^{1/p} \,.
\end{equation}
Here, $x_i$ is the i-th component of the discrete portfolio allocation vector $\vec{x}$ and $S_i$ is the ESG score of the i-th component stock in the portfolio. All other quantities are defined as before.

\begin{figure}[t!]
\includegraphics[width=\linewidth]{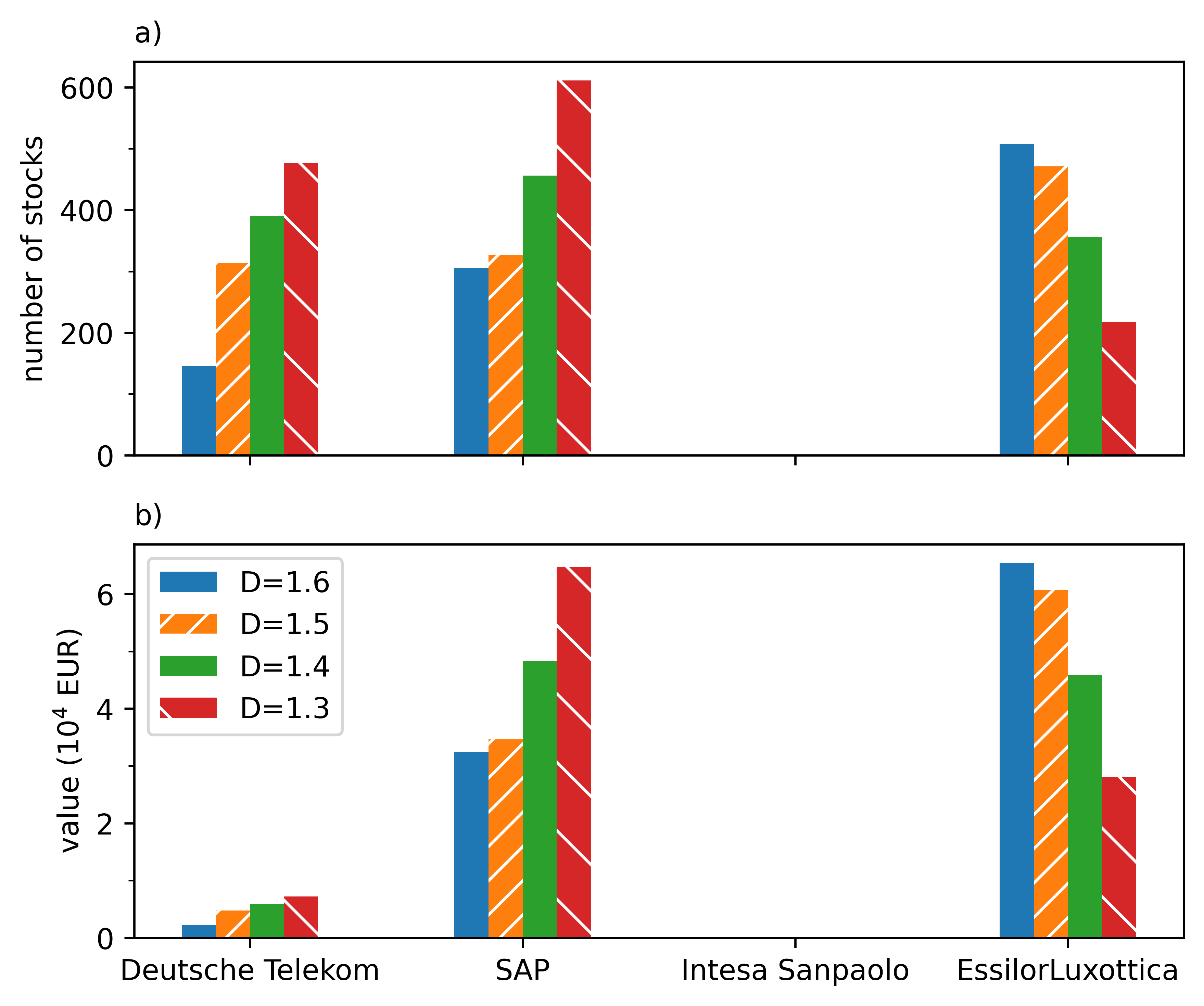}
\caption{Best discrete portfolio compositions for a budget of $B = 100000$~€, a risk-aversion parameter of $\phi=8$ and different values of maximum allowed ESG distance $D$. The ESG data were provided by ISS ESG. The grading system is on a scale from $4$ to $1$, where a higher number indicates better ESG performance. The investment universe comprises Deutsche Telekom (ISIN DE0005557508), SAP (ISIN DE0007164600), Intesa Sanpaolo (ISIN IT0000072618) and EssilorLuxottica (ISIN FR0000121667). Those are given in order from highest to lowest ESG score. Data is taken from the period between 1 January 2010 and 1 January 2021. Decreasing the maximum distance $D$ to the portfolio with best possible ESG score results in compositions which gradually contain higher amounts of stocks with better ESG scores such as Deutsche Telekom and SAP. a) Portfolio composition in terms of number of stocks. b) Portfolio composition in terms of Euro value.
}
\label{fig:portfolioCompositionESG}
\end{figure}

The Wasserstein $p$-distance is defined for $p\in[1,+\infty)$. Since the function $f(x) = x^p$ is strictly increasing for $p \geq 1$ and $x > 0$, we know that $D_\text{ESG}(p, \vec{x})$ from eq.~\ref{eq:wassersteinstocks} has a global maximum equal to $D_{\max} = \left| S^+- S^- \right|$. Therefore, we can include a linear constraint on $D_\text{ESG}(p, \vec{x})$ into the discrete optimization problem from subsection~\ref{sec:DMPTBudget} for every $p \geq 1$. The respective problem then reads:
\begin{equation}
\label{eq:budget_utility_esg}
\min \left\{ Q_d(\vec{r}, \Sigma,\phi_c,N_\text{tot}) \right\} = \min \left\{ \frac{\phi_c}{2 N_\text{tot}} \vec{x}^T \Sigma \vec{x} - \vec{r}^T \vec{x} \right\}\,.
\end{equation}
The minimization of this utility function is subject to the following constraints:
\begin{equation}
\begin{split}
\label{eq:utilitybudgetconstraintsesg}
&\vec{x}=(x_1, \dots, x_k) \quad \text{with} \quad x_i \in \mathbb{N} \quad \forall i,\\
&\sum_{i=1}^k x_i  = N_\text{tot},\\
&\vec{p}^T \vec{x} - B \leq 0 , \\
&D_\text{ESG}(p, \vec{x}) \leq D \,.
\end{split}
\end{equation}
Note how the utility function in eq.~\ref{eq:budget_utility_esg} is unchanged compared to eq.~\ref{eq:dmptutility} and eq.~\ref{eq:budget_utility}. The difference lies only in the additional constraint in eq.~\ref{eq:utilitybudgetconstraintsesg}. Here, $D$ is a non-negative constant. For $D \geq D_{\max}$ this constraint has no effect on the optimal portfolio composition $\vec{x}$. For $D=0$ only stocks with the best possible score are allowed. In between these two extremes, the constraint restricts possible solutions to the given maximum distance in ESG rating space. In practice, we use $p = 1$ and the latest ESG date in the period under investigation to calculate $D_\text{ESG}(p = 1, \vec{x})$. Exploring the effect of other choices for $p$ is left for future studies.

We now perform calculations with the following stock universe reported in order from highest to lowest ESG score: Deutsche Telekom (ISIN DE0005557508), SAP (ISIN DE0007164600),
Intesa Sanpaolo (ISIN IT0000072618) as well as EssilorLuxottica (ISIN FR0000121667). The portfolio optimization problem from eqs.~\ref{eq:budget_utility_esg} and \ref{eq:utilitybudgetconstraintsesg} was again solved on a D-Wave quantum annealer for different values of the ESG constraint $D$. The ESG data were provided by ISS ESG. The grading system is on a scale from $4$ to $1$, where a higher number indicates better ESG performance. We use a budget of $B = 100000$~€ and a risk-aversion parameter of $\phi=8$. The result in volatility-return space is shown in Fig.~\ref{fig:positionESGfrontier}. We first set $D=5$ and obtained a solution with $D_\text{ESG} = 1.6$. Hence we conclude that the actual maximum reachable ESG distance within the given investment universe is $D_\text{ESG} = 1.6$. We gradually decreased $D$ from there until the solution visibly departed from the efficient frontier. The latter was again calculated by sampling random portfolio compositions. At a certain point, stronger constraints on $D_\text{ESG}(p, \vec{x})$ produce portfolios, which move farther away from the efficient frontier. This finding is consistent with the study by Cesarone \textit{et al.}~\citep{Cesarone2022}

We also analyzed the portfolio composition for different values of $D$. The results are shown in Fig.~\ref{fig:portfolioCompositionESG}. We found that decreasing the distance $D$ from the best possible portfolio in ESG terms gradually increases the weight of stocks with better ESG score compared to stocks with worse ESG score, both in terms of relative composition and budget allocation. Stocks of Intesa Sanpaolo are not part of the optimal portfolios due to their relatively unfavorable returns (not driven by ESG scores). We had to vary $N_\text{tot}$ somewhat as a function of $D$, so that the full budget can be allocated. As can be seen from Fig.~\ref{fig:portfolioCompositionESG}, allocation to Deutsche Telekom increases with decreasing $D$. Since stocks of Deutsche Telekom have a much lower price per stock than SAP and EssilorLuxottica, a higher number of stocks has to be allocated, which requires larger $N_\text{tot}$. The resulting budget allocations are summarized in Table~\ref{tab:budget}.

\begin{table}[t]
    \centering
    \begin{tabular}{r|r|r}
        $D$ & $N_\text{tot}$ & $\vec{p}^T \vec{x}$ in €\\
        \hline
        1.6 & 960 & 99999.56 \\
        1.5 & 1112 & 99998.61 \\
        1.4 & 1202 & 99994.84 \\
        1.3 & 1305 & 99932.86 \\
    \end{tabular}
    \caption{Number of stocks and allocated budget $\vec{p}^T \vec{x}$ as a function of the maximum ESG distance $D$.}
    \label{tab:budget}
\end{table}

In this subsection, we have introduced a novel distance measure for portfolios within the space of ESG scores based on the Wasserstein metric. We use this distance measure to constrain the search for optimal portfolios in volatility-return space to a certain vicinity of the best possible portfolio in ESG space. We have demonstrated that our approach yields sensible and interesting results in combination with discrete portfolio optimization on a quantum annealer.

\section{Discussion}
\label{sec:discussion}
The approach we have presented here is based on historical data for covariance and returns. A further constraint such as the one on the ESG classification may not improve the performance of any portfolio within this framework. However, there is an ongoing discussion in the literature on whether ESG-aware investors generate higher returns than comparable non-ESG benchmarks in the long-term and can realize a better performance during a global crisis. Results of investigations into the historically measured performance of stocks with strong and weak ESG ratings vary depending on markets, ESG data and time periods considered for analysis~\citep{Auer2016, Breedt2018, Cesarone2022,Nofsinger2014,LaTorre2020,Amon2021, Bae2021, Demers2021, Atz2023, Garcia2022}. 

Cesarone \textit{et al.}~\citep{Cesarone2022} investigate mean-variance-ESG optimal portfolios and show how portfolio mean returns systematically move away from the efficient frontier the more weight is put optimizing the ESG scores of the respective portfolio (compare Fig.~\ref{fig:positionESGfrontier}). These authors use a continuous Markowitz framework and obtain results consistent with ours. In addition, we show that optimal discrete portfolios can be obtained from modern quantum annealers under realistic circumstances.

Auer and Schuhmacher~\citep{Auer2016} study the impact of socially responsible investing on the performance of investment funds. They compare the returns of investment funds with different ESG ratings to the return of their respective benchmark index. They find that portfolios of European stocks with high ESG ratings often under-perform with respect to their benchmark, while no consistent over- or under-performance was observed in the Asia-Pacific region and the United States. This approach differs from ours in that Auer and Schuhmacher use benchmark indices as their reference point, while we compare to portfolios on the mean-variance efficient frontier. In this sense, an over-performance of ESG-aware portfolios is possible in Auer and Schuhmacher's approach, since they compare to benchmark indices, which may have sub-optimal returns in the first place. Due to the different methodology, these authors' results are not directly comparable to ours. Nevertheless, our results and those of ref.~\onlinecite{Cesarone2022} can help to rationalize these findings. Both studies find that the deviation of ESG-aware from mean-variance optimal portfolios depends on the emphasis, which is put on the ESG optimization goal. In particular, the novel ESG distance measure we introduced could help to clarify the results of Auer and Schuhmacher in future studies.

Amon \textit{et al.}~\citep{Amon2021} find that portfolios with good ESG ratings can be constructed at a small cost in terms of returns. This is consistent with our findings and those of ref.~\onlinecite{Cesarone2022}, which both show that many portfolios with close to optimal returns can at the same time have good ESG ratings. In future studies, our ESG distance measure could be used to quantify the deviation of these ESG-aware portfolios from the best possible portfolio in the respective rating system.

Garc\'ia \textit{et al.}~\citep{Garcia2022} investigate ESG ratings within a multi-objective optimization framework, focusing on portfolios comprised of component stocks from the Dow Jones Industrial Average (DJIA) index. These authors solve an NP-hard realistic portfolio problem similar to ours, but use a heuristic evolutionary algorithm where we employ a quantum annealer. They find that better ESG ratings generally imply lower returns. Nevertheless, many portfolios with good ESG ratings possess favorable risk-return profiles and may even outperform benchmark indices like the DJIA. These results are consistent with our present study.

Breedt \textit{et al.}~\citep{Breedt2018} perform a factor analysis based on a proprietary mean-variance optimization method. They find that ESG is not an independent factor, i.e. ESG information is already captured by other investment factors. They conclude that including ESG information into the investment process does neither lower nor improve the investment returns. We found that it is possible to construct ESG-aware portfolios, which are very close to the efficient frontier. Hence, our results can be considered consistent with those of Breedt and coauthors.

Nofsinger and Varma~\citep{Nofsinger2014} find that socially responsible investment portfolios over-perform in times of market crisis and under-perform in other periods. They performed regression using several factor models. This methodology is very different from ours and other mean-variance approaches. Furthermore, the ESG selection is based on a screening approach, not on optimization. Again, over- and under-performance was measured with respect to regional benchmarks. Therefore, these results are not directly comparable to ours.

Demers \textit{et al.}~\citep{Demers2021} performed a similar study and concluded that ESG-aware investment does not protect against market crises. Their argument is similar to that of ref.~\onlinecite{Breedt2018}, since they also conclude that ESG is not an independent investment factor.

Bae \textit{et al.}~\citep{Bae2021} perform a regression analysis and conclude that corporate social responsibility does not affect the returns of US stocks during times of the COVID-19 market crisis. These authors also point out the possibility of firms having positive ESG values in certain rating systems, while actually acting against these goals in practice. Like other factor regression studies, these results are not directly comparable to ours.

La Torre \textit{et al.}~\citep{LaTorre2020} find that ESG ratings weakly affect the returns of EURO STOXX 50 component stocks. Their analysis is based on regression of a factor model, which is only loosely related to our study. Again, our quantitative distance measure in ESG space could help to clarify these results in future studies.

Atz \textit{et al.}~\citep{Atz2023} perform a meta-study on the impact of sustainability on investment returns. They argue that most studies find no discernible impact, while about one-third of all investigated studies find a positive impact. The positive impact is attributed to the possibility of capturing climate risk premium and higher robustness during times of crisis.

As explained before, the question of whether ESG-aware investing produces measurable effects on investment performance is beyond the scope of this work. Such effects would result from investment decisions guided by beliefs and values, which are not captured by the Markowitz framework used in our study.

In our opinion, the question is ultimately to which degree investor expectations about future developments are reflected in historical prices. As we explained in the introduction, the importance of informed investment decisions based on ESG data can be expected to grow. The degree to which non-ESG-aware investors are following these developments is, however, unclear. Therefore, stocks with good ESG scores may outperform stocks with worse ESG scores, as the public increasingly demands the publication of ESG data and enforces the adoption of ESG-aware investment strategies. This effect is not captured by Markowitz portfolio theory and would require a radically different approach.

We expect that interest in ESG topics will grow rapidly among investors, particularly regarding portfolio classification in terms of ESG scores. Our method of using ESG data enables ESG-aware investors to construct ESG-friendly portfolios without the need for further assumptions or additional parameters. In particular, we avoid assuming additivity of ESG data with other terms in the Markowitz utility function. In fact, we do not modify the utility function at all, so that ESG data only appears in the linear constraint we introduced. Thus, in our approach ESG preference, returns and volatility can be interpreted independently, as one would expect~\citep{Pedersen2021,Varmaz2022,Utz2014}.

Our study also shows that portfolio optimization is an attractive case for combining classical and quantum workflows. While the discrete portfolio optimization problem can only be solved efficiently on a quantum computer, all data processing is still done efficiently on a classical computer. We believe that many quantum applications will be part of such hybrid quantum-classical workflows in the future. See refs.~\onlinecite{Sakuler2023,Cohen2020,Cohen2020_60Stocks,Mugel2021,Venturelli2019,Lang2022} for more examples for hybrid approaches to portfolio optimization.

\section{Conclusions}
\label{sec:conclusions}
We have presented a study of Markowitz portfolio optimization in the presence of discrete stock allocations, limited budget and constraints on portfolio ESG scores. We have studied both the usual continuous formulation of the portfolio problem as well as a more realistic discrete version. The discrete version can not be solved efficiently on classical computers, at least not by enumerating all possible portfolio combinations, although some progress has been achieved using simulated annealing on classical hardware~\citep{Rubio2022}. Therefore, we have employed a D-Wave quantum annealer for solving the discrete portfolio problem.

We have established a mapping between continuous and discrete Markowitz portfolio theory, which allows us to compare results in a meaningful way. This mapping involves a re-scaling of the risk-aversion parameter $\phi$. Importantly, we have also shown that failing to apply this re-scaling in the discrete case, the relative composition of discrete solutions will not converge to the continuous solution, even in the limit of infinite portfolio size, but rather converge to the minimum variance portfolio.

Subsequently, we extended Markowitz portfolio theory to include a budget constraint. We showed that rounding of continuous portfolio compositions to the nearest integer number of stocks yields sub-optimal portfolios for small and medium investment budgets. Solutions from our discrete approach on the contrary lie on the efficient frontier in volatility-return space.

Furthermore, we introduced a novel way to classify portfolios in terms of their ESG score via the Wasserstein $p$-distance by viewing relative portfolio compositions as discrete probability distributions. Using the Wasserstein metric we measure a portfolio's distance with respect to the best possible portfolio by ESG score in the respective scoring system. Our method is a generalization of the weighted average classification scheme reported in the literature and applicable to any ESG scoring system without further modification. Our framework can even be modified to accommodate ESG data from heterogeneous scoring systems. We incorporated the ESG data into the optimization process by constraining the portfolio search to a certain maximum distance from the portfolio with the best possible ESG score via a linear constraint, which is independent of the chosen metric.

We also reported case studies for portfolios using components of the well-known EURO STOXX 50 index. By decreasing the maximum distance from the best ESG portfolio we found that portfolio compositions were gradually putting more weight on stocks with better ESG scores and less weight on stocks with worse ESG scores, both in terms of number of stocks and in terms of allocated budget.

Our results can help ESG-aware investors to include their preferences in an effective way, building on the widely used Markowitz portfolio theory. How these preferences are derived is a research field in itself and goes beyond our work.

So far, we have only studied the Wasserstein $p$-distance for $p=1$. Future studies should clarify the role of $p$ for the ESG portfolio problem. Furthermore, our method could be applied to larger portfolios and heterogeneous ESG data from different providers. We believe that the formalism we presented can be applied to many practical problems, such as finding tradeable ESG-optimized portfolios or constructing discrete ESG-aware portfolios as a basis for exactly hedgeable indices. These topics are left for future research.

\bibliography{paper_arxiv.bib}

\end{document}